\documentclass{kapproc} 

\usepackage{procps} 

\usepackage[dvips]{graphicx}

\upperandlowercase

\kluwerbib 

\def\plotfiddle#1#2#3#4#5#6#7{\centering \leavevmode
\vbox to#2{\rule{0pt}{#2}}
\includegraphics{#1}}

\begin{document}



\articletitle[Asymmetric Drift and the Stellar Velocity Ellipsoid]{Asymmetric Drift \\ and the Stellar Velocity Ellipsoid}

\chaptitlerunninghead{The Stellar Velocity Ellipsoid}















\author{Kyle B. Westfall\altaffilmark{1}, Matthew A. Bershady\altaffilmark{1},
Marc A. W. Verheijen\altaffilmark{2},\\ David R. Andersen\altaffilmark{3},
\& Rob A. Swaters\altaffilmark{4}} 

\affil{
\altaffilmark{1}University of Wisconsin -- Madison, \
\altaffilmark{2}Kapteyn Institute, \ \\
\altaffilmark{3}NRC Herzberg Institute of Astrophysics, \
\altaffilmark{4}University of Maryland
}





\begin{abstract}
We present the decomposition of the stellar velocity ellipsoid using stellar
velocity dispersions within a 40$^{\circ}$ wedge about the major-axis
($\sigma_{\rm maj}$), the epicycle approximation, and the asymmetric drift
equation.  Thus, we employ no fitted forms for $\sigma_{\rm maj}$ and escape
interpolation errors resulting from comparisons of the major and minor axes.
We apply the theoretical construction of the method to integral field data
taken for NGC 3949 and NGC 3982.  We derive the vertical-to-radial velocity
dispersion ratio ($\sigma_z/\sigma_R$) and find (1) our decomposition method
is accurate and reasonable, (2) NGC 3982 appears to be rather typical of an
Sb type galaxy with $\sigma_z/\sigma_R = 0.73^{+0.13}_{-0.11}$ despite its
high surface brightness and small size, and (3) NGC 3949 has a hot disk with
$\sigma_z/\sigma_R = 1.18^{+0.36}_{-0.28}$.
\end{abstract}



\section{Motivation and Methodology}

The shape of the stellar velocity ellipsoid, defined by $\sigma_R$,
$\sigma_{\phi}$, and $\sigma_z$, provides key insights into the
dynamical state of a galactic disk: $\sigma_z$:$\sigma_R$ provides a
measure of disk heating and $\sigma_{\phi}$:$\sigma_R$ yields a check
on the validity of the epicycle approximation (EA).  Additionally,
$\sigma_R$ is a key component in measuring the stability criterion and in
correcting rotation curves for asymmetric drift (AD), while $\sigma_z$ is
required for measuring the disk mass-to-light ratio.  The latter is
where the DiskMass survey focuses (Verheijen et al. 2004,
2005); however, in anything but face-on systems, $\sigma_z$ must
be extracted via decomposition of the line-of-sight (LOS) velocity
dispersion.
Below, we present such a decomposition for two galaxies in
the DiskMass sample: NGC 3949 and NGC 3982.

Previous long-slit studies (e.g., Shapiro et al. 2003 and references
therein) acquired observations along the major and minor axes and
performed the decomposition via the EA and AD equations; using both
dynamical equations overspecifies the problem such that AD is often used
as a consistency check.  Here, use of the SparsePak (Bershady et
al. 2004, 2005)
integral field unit (IFU) automatically provides multiple position angles,
thereby increasing observing efficiency and ensuring signal extraction
along the desired kinematic axes.  Long-slit studies
have also used functional forms to reduce the sensitivity of the above
decomposition method to noise.  Here, only measures of the LOS velocity
dispersions within a 40$^{\circ}$ wedge about the major axis are used to
perform the decomposition by
incorporating both the EA and AD equations under some simplifying
assumptions.
Velocities and radii within the wedge are projected onto the major axis
according to derived
disk inclinations, $i$, and assuming near circular motion.  In the end,
our method requires neither fitted forms nor error-prone interpolation between
the major and minor axes.  Future work will compare this decomposition method
with the multi-axis
long-slit method and investigate effects due to use of points off the kinematic
axes.
 
Following derivations in Binney \& Tremaine (1987) and
assuming (1) EA holds, (2) the velocity ellipsoid shape and orientation is
independent of $z$
($\partial(\overline{v_R v_z})/ \partial z = 0$), (3) both the space
density, $\nu$, and $\sigma_R$ have an exponential fall off radially with scale
lengths of $h_R$ and $2h_R$, respectively, and (4) the circular velocity
is well-represented by the gaseous velocity, $v_g$, the equation for the AD
of the stars becomes
%
$v^2_g - \overline{v_{\ast}}^2 =  \frac{1}{2} \sigma^2_R \left( \frac{\partial \mbox{ln}\overline{v_{\ast}}}{\partial\mbox{ln}R} + 4 \frac{R}{h_R} - 1 \right)$,
%
where $\overline{v_{\ast}}$ is the mean stellar rotation velocity; hence,
$\sigma_R$ is the only unknown.  The third assumption requires mass to follow
light, $\Sigma \propto \sigma_z^2$, and constant velocity ellipsoid axis ratios
with radius; $\Sigma$ is the surface density.  The major-axis dispersion is
geometrically given by $\sigma_{maj}^2 = \sigma_{\phi}^2 {\rm sin}^2i +
(\eta\sigma_R)^2{\rm cos}^2i$, where $\eta=\sigma_z/\sigma_R$ is constant
with radius.  Finally, EA, $\sigma^2_{\phi}/\sigma^2_R = \frac{1}{2} \left(\frac{\partial \mbox{ln}\overline{v_{\ast}}}{\partial \mbox{ln}R} + 1\right)$,
completes a full set of equations for decomposition of the velocity ellipsoid.

\section{Analysis}

Data for testing of the above formalism was obtained during SparsePak
commissioning (Bershady et al. 2005, see Table 1).
Both NGC 3949 and NGC 3982 were observed for $3\times2700$s at one IFU
position with $\lambda_c=513.1$ nm and $\lambda/\Delta\lambda = 11700$ or
26 km s$^{-1}$.
The velocity distribution function (VDF) of both gas and stars is
parameterized by a Gaussian function.  In each fiber with sufficient
signal-to-noise, the gaseous VDF is extracted using fits to the
[{\scshape Oiii}] emission line; the stellar VDF is extracted using a modified
cross-correlation method (Tonry \& Davis 1979; Statler 1995)
with HR 7615 (K0III) as the template (Westfall et al. 2005).  The pointing
of the IFU on the galaxy is determined {\it post factum} to better than 1"
by minimizing the
$\chi^2$ difference between the fiber continuum flux and the surface brightness
profile.  Subsequent galactic coordinates have been deprojected according to
the kinematic $i$ and position angle.

Figure 1 shows LOS dispersions for both the gas, $\sigma_{g, LOS}$, and stars,
$\sigma_{\ast, LOS}$; data points are given across the full field of the IFU
with points along the major and minor axes and in between having different
symbols (see caption).  From this Figure note (1) there is no significant
difference in $\sigma_{LOS}$ along the major and minor axes for NGC 3949 and
(2) the large gas dispersion within $R\leq 10$" for NGC 3982 is a result of
poor single Gaussian fits to the multiple dynamical components of its LINER
nucleus.
\begin{figure}[ht]
\plotfiddle{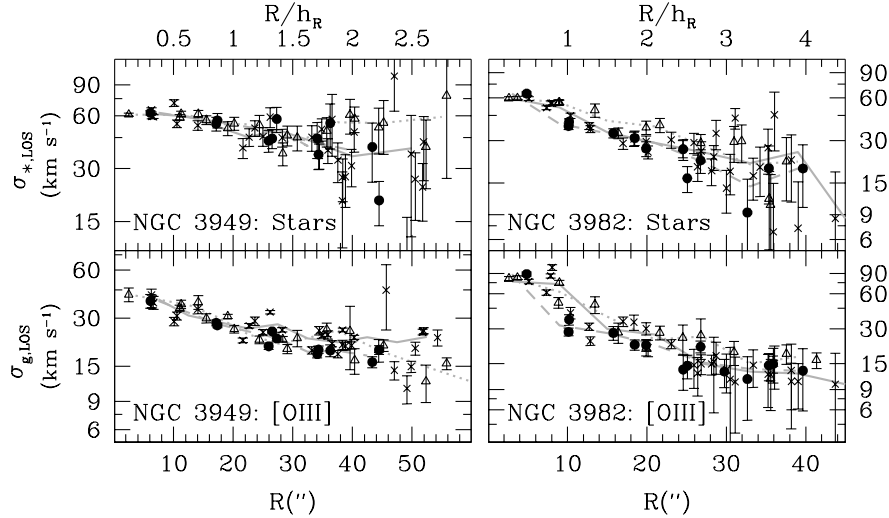}{2.2in}{0}{60}{60}{-180}{-107}
\caption{ Measurements of the LOS dispersions in NGC 3949 ({\em left}) and
  NGC 3982 ({\em right}) for stars ({\em top}) and gas ({\em bottom}) as a
  function of galactic radius; the top axis is in units of scale length
  ($h_R$ is 20 and 10 arcsec in NGC 3949 and NGC 3982, respectively).
  Major axis, minor axis, and in between fibers are shown as
  points, triangles, and crosses, respectively; ``axis fibers'' lie in a
  $40^{\circ}$ wedge around the axis.  The binned mean values are
  shown for the major axis, minor axis, and otherwise as dashed, dotted, and
  solid lines, respectively.
}
\end{figure}

Figure 2 gives the folded gaseous and stellar rotation curves and compares
the measured $\sigma_{\rm maj}$ from Figure 1 with that
calculated using the formalism from \S1.  The value of $\eta$ used in
Figure 2 provides the minimum difference between the two sets of data (as
measured by $\chi^2_{\nu}$; see Figure 3a).  We find
$\eta = 1.18^{+0.36}_{-0.28}$ and $\eta = 0.73^{+0.13}_{-0.11}$
for NGC 3949 and NGC 3982, respectively; errors are given by 68\%
confidence limits.  A comparison of these values to the summary in
Shapiro et al. (2003) is shown in Figure 3b.  The disk of NGC
3982 is similar to other Sb types studied; however, NGC 3949 seems to
have an inordinately hot disk.  The latter, while peculiar, is also
supported by 
the indifference between its major and minor axis $\sigma_{LOS}$ from Figure 1
and the same indifference seen in the Ca{\scshape ii} data presented by
Bershady et al. (2002).  Our streamlined velocity-ellipsoid decomposition
method appears accurate, as seen by comparison with (1) galaxies of a similar
type for NGC 3982, and (2) data in a different spectral region for NGC
3949.\newline

\noindent This work was generously supported by an AAS International Travel
Grant, the Wisconsin Space Grant Consortium, and the NSF Grant AST-0307417.
\begin{figure}[ht]
\plotfiddle{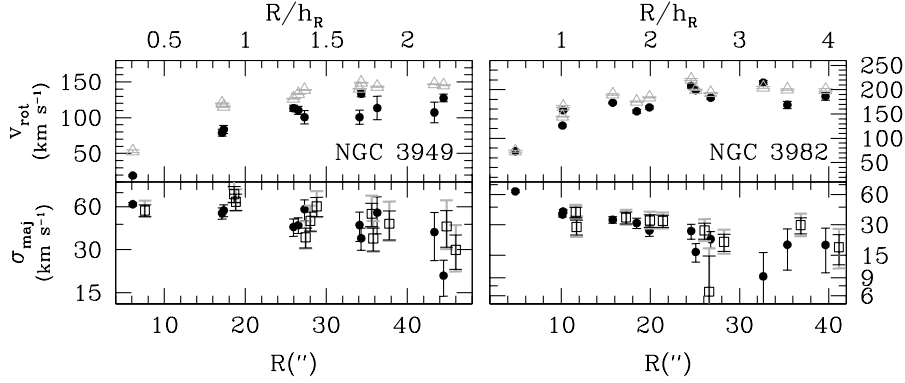}{1.7in}{0}{60}{60}{-180}{-160}
\caption{Comparison of measured and calculated $\sigma_{maj}$ in
  NGC 3949 ({\em left}) and NGC 3982 ({\em right}). Top panels show
  the deprojected rotational velocity for both the gas
  ({\em grey triangles}) and stars ({\em black points}) along the major axis.
  Bottom panels compare the measured $\sigma_{maj}$ ({\em points}) to those
  from our formalism using $\eta$ as derived from Figure 3a ({\em squares};
  offset in $R$ for comparison).  Errors on $\sigma_{maj}$ are also calculated
  when using the 68\% confidence limits for $\eta$ ({\em grey}). }
\end{figure}
\begin{figure}[ht]
\plotfiddle{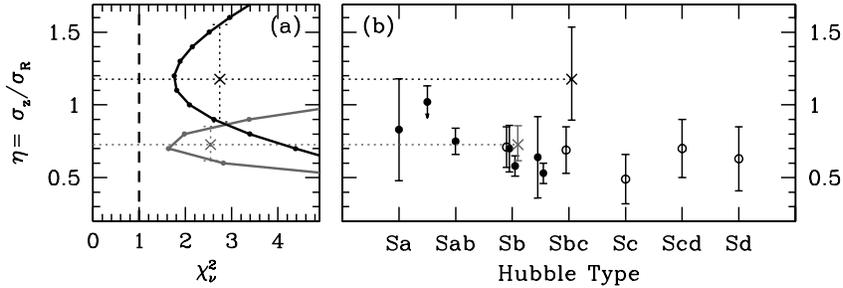}{1.0in}{0}{55}{55}{-165}{-190}
\caption{Derivation of the ratio $\eta=\sigma_z/\sigma_R$.  In (a), the
  $\chi^2_{\nu}$ statistic is used to find the best correspondence between the
  model and measured data sets for NGC 3949 ({\em black line}) and NGC 3982
  ({\em grey line}).  The 68\% (1$\sigma$) confidence limits on $\eta$ are
  determined by an increase of $\chi^2_{\nu}$ by 1 ({\em dotted lines}).  The
  disk heating of NGC 3949 ({\em black cross}) and NGC 3982 ({\em grey cross})
  are placed on Figure 5 from Shapiro et al. (2003) in (b).}
\end{figure}






%




\end{document}